\documentclass[lettersize,journal]{IEEEtran}

\IEEEoverridecommandlockouts
\renewcommand{\baselinestretch}{0.9721}

\usepackage{cite}
\usepackage{algorithmic}
\usepackage[linesnumbered,ruled,vlined]{algorithm2e}
\algsetup{linenosize=\small}
\usepackage{graphicx}
\usepackage{multirow}
\usepackage{textcomp}
\usepackage{xcolor}
\usepackage{scalerel}
\usepackage{graphics}
\usepackage[utf8]{inputenc}
\usepackage{amsfonts}
\usepackage[fleqn]{amsmath}
\usepackage{tikz}
\usepackage[hidelinks]{hyperref}
\usepackage{empheq}
\usepackage{array}
\usepackage{multirow}
\usepackage{booktabs}
\usepackage{amssymb}
\usepackage{color}
\usepackage{array}
\usepackage{hhline}
\usepackage{soul}
\usepackage{comment}
\usepackage{tablefootnote}
\usepackage{stfloats}
\usetikzlibrary{svg.path}

\definecolor{orcidlogocol}{HTML}{A6CE39}
\tikzset{
    orcidlogo/.pic={
        \fill[orcidlogocol] svg{M256,128c0,70.7-57.3,128-128,128C57.3,256,0,198.7,0,128C0,57.3,57.3,0,128,0C198.7,0,256,57.3,256,128z};
        \fill[white] svg{M86.3,186.2H70.9V79.1h15.4v48.4V186.2z}
        svg{M108.9,79.1h41.6c39.6,0,57,28.3,57,53.6c0,27.5-21.5,53.6-56.8,53.6h-41.8V79.1z M124.3,172.4h24.5c34.9,0,42.9-26.5,42.9-39.7c0-21.5-13.7-39.7-43.7-39.7h-23.7V172.4z}
        svg{M88.7,56.8c0,5.5-4.5,10.1-10.1,10.1c-5.6,0-10.1-4.6-10.1-10.1c0-5.6,4.5-10.1,10.1-10.1C84.2,46.7,88.7,51.3,88.7,56.8z};
    }
}

\newcommand\orcidicon[1]{\href{https://orcid.org/#1}{\mbox{\scalerel*{
                \begin{tikzpicture}[yscale=-1,transform shape]
                \pic{orcidlogo};
                \end{tikzpicture}
            }{|}}}}

\def\BibTeX{{\rm B\kern-.05em{\sc i\kern-.025em b}\kern-.08em
    T\kern-.1667em\lower.7ex\hbox{E}\kern-.125emX}}
    
    \newcommand{\etal}{\textit{et al. }}
    
\begin{document}

\title{Hardware Trojan Insertion in Finalized Layouts:\\From Methodology to a Silicon Demonstration\\[-0.1cm]}

\author{Tiago Perez\textsuperscript{\orcidicon{0000-0001-6006-1938}}, Samuel Pagliarini\textsuperscript{\orcidicon{0000-0002-5294-0606}}, Member,~IEEE
\thanks{Manuscript received June, 2022. \\This work has been partially conducted in the project ``ICT~programme'' which was supported by the European Union through the European Social Fund. It was also partially supported by the Estonian Research Council grant MOBERC35. 

Tiago Perez and Samuel Pagliariani are with the the Department of Computer Systems, Tallinn University of Technology, Tallinn, Estonia (email: tiago.perez@taltech.ee; samuel.pagliarini@taltech.ee).
}}

\markboth{IEEE TRANSACTIONS ON COMPUTER-AIDED DESIGN OF INTEGRATED CIRCUITS AND SYSTEMS, VOL. xx, NO. x, XXXX 2022}%
{How to Use the IEEEtran \LaTeX \ Templates}

\maketitle

\begin{abstract}
Owning a high-end semiconductor foundry is a luxury very few companies can afford. Thus, fabless design companies outsource integrated circuit fabrication to third parties. Within foundries, rogue elements may gain access to the customer's layout and perform malicious acts, including the insertion of a hardware trojan (HT). Many works focus on the structure/effects of a HT, while very few have demonstrated the viability of their HTs in silicon. Even fewer disclose how HTs are inserted or the time required for this activity. Our work details, for the first time, how effortlessly a HT can be inserted into a finalized layout by presenting an insertion framework based on the engineering change order flow. For validation, we have built an ASIC prototype in 65nm CMOS technology comprising of four trojaned cryptocores. A side-channel HT is inserted in each core with the intent of leaking the cryptokey over a power channel. Moreover, we have determined that the entire attack can be mounted in a little over one hour. We also show that the attack was successful for all tested samples. Finally, our measurements demonstrate the robustness of our SCT against skews in the manufacturing process.
\end{abstract}


\section{Introduction} \label{sec:introduction}

  
  \IEEEPARstart{T}{he} ever-increasing cost to build high-end semiconductor manufacturing facilities -- building a 3nm production line is estimated to cost \$15-20B \cite{Cost3nm} -- has made most design companies migrate to a fabless business model. In practice, fabless design houses can design and market integrated circuit (IC) solutions, but the actual IC fabrication is outsourced to a third party. This practice can potentially affect the trustworthiness of an IC as a foundry (or a \emph{rogue element} within the foundry) can manipulate the design for malicious purposes \cite{Guin2014}. Many fabrication-time threats have been studied recently \cite{outs_threats}. For combating these threats, numerous techniques have been proposed for increasing the trustworthiness of an IC. Examples of these techniques are Split Manufacturing \cite{split_survey}, Logic Locking \cite{logic_1, logic_3, new_logic_1} and IC Camouflaging \cite{cam_3}. Unfortunately, the current state of these techniques makes them unsuitable for large-scale production of ICs, either because of practicality \cite{split_survey} and/or insufficient security guarantees \cite{eval_logic}. Thus, the current practices of a globalized supply chain leave the fabrication of ICs vulnerable to attacks.
  
  \begin{figure*}[tb]
    \centering
    \includegraphics[width=0.9\linewidth]{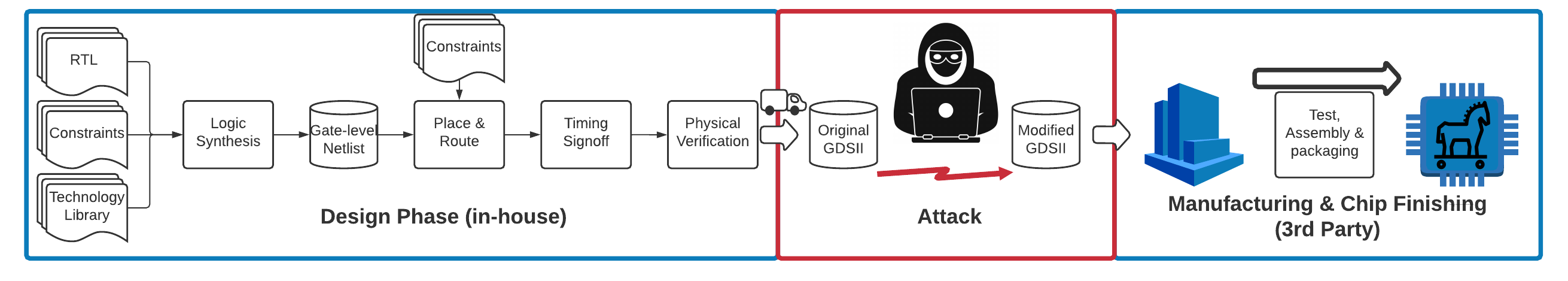}
    \caption{A typical IC design flow. Highlighted in red is the stage where a rogue element may mount an attack (modified from \cite{trojan_iscas}).}
    \label{fig:phys_imp_flow}
\end{figure*}
  
  Tampering an otherwise trustworthy IC can be done by inserting malicious logic or modifying specific aspects of the manufacturing process \cite{Tehranipoor2010,Rostami2014}. These kinds of modifications are often referred to as hardware trojans (HTs). HTs are designed to leak confidential information, to disrupt a system's specific functionality, or even to destroy the entire system (referred to as time-bomb). Various types of HTs have been studied recently \cite{MolesSDT, LightTrojan, WLTrojan, HT_silicon, ParaFaultInj,A2AnaTrojan, AmpSDLCrypto, Ali2021, sct_param}, demonstrating the potential threat of this type of attack. 
  
  An IC's operating physical characteristics, such as timing, power consumption, electromagnetic radiation, and even sound, can be used as a side-channel to indirectly reveal information that should be internal to the IC. For this reason, side-channel attacks (SCAs) often target keys of embedded crypto cores \cite{FirstDPA}. However, to mount a successful SCA, acquiring a large amount of data is usually required, followed by correlation/statistical analysis. Moreover, a very specific type of HT has been proposed for assisting SCAs. Lin \etal \cite{MolesSDT} were the first to propose an HT architecture for assisting a power SCA, referred to as ``Malicious Off-chip Leakage Enabled by Side-channels'' (MOLES). This specific type of trojan is the centerpiece of our work; in the remainder of this text it is referred to as a side-channel trojan (SCT). By using SCTs, the attack time can be drastically reduced as no further processing is required. The disadvantage of SCTs is their invasive nature. Inserting an SCT requires a modification of the circuit at fabrication time. While this might seem a difficult task at first sight, we later show how it can be executed by a capable attacker.

Despite encouraging results reported from the SCT studies, only in \cite{trojan_iscas} the authors discuss how SCTs could be inserted \textbf{from the perspective of the attacker}. Even more concerning, this discussion is also absent from silicon-validated trojans \cite{HT_silicon, WLTrojan, sct_param}. In this work, we present an extension of the SCT design methodology described in \cite{trojan_iscas}. We assume that a rogue element inside the foundry is the adversary and that he/she makes use of \textbf{readily available engineering change order (ECO)} capabilities of physical design tools. Consequently, the main contribution of this work, in addition to a framework for inserting SCTs, is an ASIC prototype for validating our methodology. A rich discussion regarding the effectiveness of the approach is also provided.

Furthermore, we have fabricated a testchip comprising 4 cryptocores in a 65nm commercial technology, validating our technique in silicon. Our in-depth analysis of the side-channel reading versus the variation in the manufacturing process demonstrated that our SCT is robust against process variation. On top of that, we include an analysis of the attack time required for inserting our SCT in a finalized layout when also utilizing our ECO-based methodology. These characteristics sharply contrast our work with the prior art.

  In the next section, we describe our considered threat model and the attacker capabilities. Our SCT architecture and our insertion methodology are detailed in Section~\ref{sec:trojan_fw}. In the Section~\ref{sec:results}, we present our experimental results. In the first part of Section~\ref{sec:results}, we assume the role of the victim by implementing the target circuits. Later, we assume the role of the adversary by designing and inserting the SCT, utilizing the methodology described in Section~\ref{sec:trojan_fw}. In Section~\ref{sec:testchip}, we present our prototype and describe the experiments executed during its validation. Our results discussion, along with a thorough discussion regarding hardware trojan detection, are presented in Section~\ref{sec:discussion}. Finally, we draw our conclusion in Section~\ref{sec:conclusion}.
  
\section{Threat Model and Attacker Capabilities} \label{sec:background}

   In this work, the principal attacker we are concerned with is a \textbf{rogue element} inside the foundry. His/her aim is to insert malicious logic into a finalized layout. We emphasize that the attack occurs before the fabrication, and a single rogue element inside the foundry is sufficient to perform the proposed attack. Thus, since the attacker is located inside the foundry, he/she enjoys access to all technology and cell libraries\footnote{This is particularly true for advanced nodes where only a handful of cell libraries per node exist. Typically, the foundry or a company licensed by the foundry provides a standard cell library. In either case, we assume the attacker has no difficulty identifying individual gates and their functionality.} utilized by the victim when creating the layout. We assume the attacker can identify the presence of a crypto core in a layout, which is a reasonable assumption, especially for well-known AES implementations that display regularity (due to the round-based key schedule structure). To be very clear, we do not assume that the adversary understands the entire victim's design (nor is there a need for such knowledge). Instead, we assume that the adversary can recognize the layout/structure of a single crypto core within a larger design, in line with the assumptions made in \cite{LightTrojan,HT_silicon}.

  Furthermore, we also assume the adversary: 1) is versed in IC design, 2) enjoys access to modern EDA tools, 3) has no means to make radical modifications to the circuit (e.g., adding new IOs or making changes in the clock domains). With the help of the inserted logic in the form of an SCT, the attacker will then attempt to leak confidential information via a power signature. Crypto cores are often the target in this type of attack \cite{HT_silicon,ParaFaultInj} -- this is also the case in our work. As our attack deals with power signature reading, stopping some part of the clock delivery, or even entirely, would be highly beneficial for the attack. However, the attacker is assumed to have no knowledge about the clock domains or clock distribution in general. Synchronizing and controlling the HT's trigger to totally stop the clock delivery is not an option we have considered feasible, nor is the addition of an external trigger controlled by an IO.

  A typical IC physical implementation flow is described in the left portion of Fig. \ref{fig:phys_imp_flow}. The attack takes place after the victim's layout in GDSII format is sent for fabrication (see red portion of Fig. \ref{fig:phys_imp_flow}). Suppose the attacker had access to all of the victims' data required to generate the layout (i.e., RTL, netlists, constraints, etc.). In this case, he/she could replicate the physical implementation flow to achieve a layout similar to the one created by the victim, yet now containing his malicious logic. This effort is theoretically possible but largely unpractical. Our threat model, therefore, assumes that the attacker only has access to the finalized layout. Design companies have to hand over their finalized layout to the foundry for fabrication. Normally, the layouts require some pre-processing steps before the start of the fabrication, which are handled by a foundry employee. Thus, it is during this time period that the attack can be mounted.
  
  

  Nevertheless, EDA tools already have the capability to deal with finalized designs. This functionality is a feature referred to as ECO. Thus, an attacker holding only the layout could use an ECO to modify or insert additional logic in a finalized layout. An ECO flow requires four inputs: a technology library, a cell library, the gate-level netlist, and a timing constraint. The adversary already possesses the first two, but must generate the third and estimate the fourth input. A gate-level netlist can be extracted from the victim's layout \cite{gds_to_gnetlist}, while the timing constraint can be estimated to a certain degree. Our proposed trojan insertion framework is shown in Fig. \ref{fig:trojan_ins}, where these two steps are considered. The details of the framework are described in the next section.
 
\section{Side-Channel Trojan Design and Insertion} \label{sec:trojan_fw}

\subsection{Side-Channel Trojan Design} \label{sec:trojan_design}
 
  Our SCT is an additive hardware trojan with the intention of aiding a side-channel attack. For more details regarding hardware trojans taxonomy and concepts we direct the reader to \cite{new_survey}. Thus, it is designed for creating an artificial power consumption through which information is leaked. This has to be performed in a controlled manner, naturally. Knowing that the majority of the power consumption in a circuit comes from the switching activity (dynamic power), a great candidate to be a controlled power sink is a structure with a controllable frequency of operation. 
  
\begin{figure}[t!]
    \centering
    \includegraphics[width=0.98\linewidth]{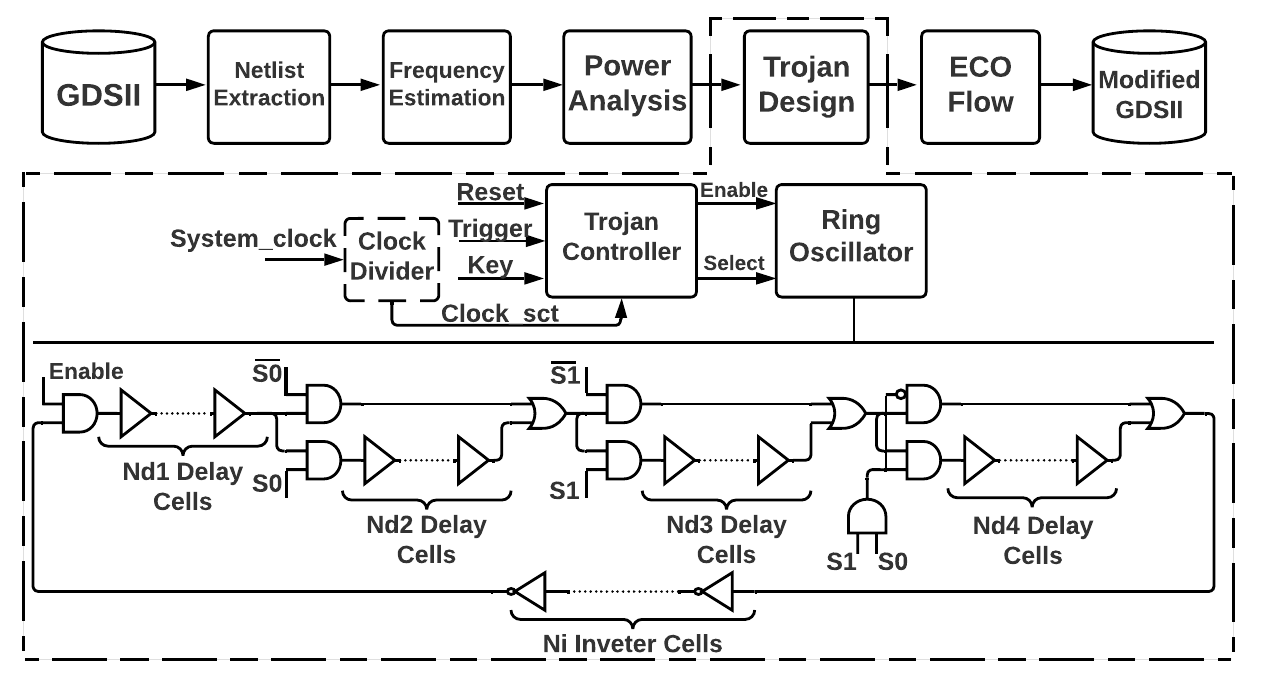}
    \caption{Our trojan insertion methodology for an SCT capable of leaking 2 bits per power signature reading (modified from \cite{trojan_iscas}).}
    \label{fig:trojan_ins}
\end{figure}

  A ring-oscillator (RO) is an example of such power sink if the number of stages in the RO can be adjusted dynamically as shown in Fig. \ref{fig:trojan_ins}. Our architecture implements variable delay stages broken into branches that are controlled by $N_{leak}$ leaking bits. Each branch of our RO has two active path options: a direct connection to the next branch or a series of delay cells. Thus, each set of the $N_{leak}$ is associated with a distinct change in the power consumption amplitude. This artificial power consumption created by the RO is similar to a pulse-amplitude modulation technique, with an order equal to $2^{N_{leak}}$. An example of this RO architecture for $N_{leak}=2$ is illustrated in Fig. \ref{fig:trojan_ins}. The active paths' configuration is described in Table \ref{tab:rosca_op_mode}, where the leaking bits become branch selectors and are referred to as S0 and S1. 

 \begin{table}[htb]
        \centering
        \caption{Ring oscillator active path configuration}
        \begin{tabular}{p{.2cm}p{.2cm}ccc}
        \hline 
          \textbf{S0} & \textbf{S1} & \textbf{Delay Cells} & \textbf{Inverter Cells} & \textbf{Freq.}\\
           \hline 
           0 & 0  & $N_{D1}$ & $N_i$ & High \\
           1 & 0  & $N_{D1}+N_{D2}$ & $N_i$ & Mid-high \\
           0 & 1  & $N_{D1}+N_{D3}$ & $N_i$ & Mid-low\\
           1 & 1  & $N_{D1}+N_{D2}+N_{D3}+N_{D4}$ & $N_i$ & Low \\
            \hline 
        \end{tabular}
        \label{tab:rosca_op_mode}
    \end{table}
    
  A dual-sided constraint guides the attacker's effort: he/she has to induce a discernible amount of dynamic power (i.e., to increase the effectiveness of the attack) while increasing leakage power as little as possible (i.e., to avoid detection). In this sense, not only the RO-based SCT structure has to be carefully planned, but a decision has to be made as to when exactly will the trojan be triggered. Our approach is to not allow the trojan to compete with the dynamic power consumption of the crypto core. Therefore, when the core is actively working, the trojan is silent and the RO is not switching. When the crypto core is idle, the trojan is triggered. For this reason, our proposed SCT trojan has a Trigger signal that is connected to the ``done'' signal coming from the crypto core, which marks the end of a cryptographic operation. 
 
  When triggered, the SCT connects a set of the leaking bits per clock cycle in the RO until all the $N_{key}$ bits from the crypto key are leaked. Thus, our SCT requires a connection to the system clock and reset, a trigger signal, and the crypto key. Its architecture is illustrated in Fig.~\ref{fig:trojan_ins}, consisting of three blocks: clock divider (DV), the trojan controller (TC), and the RO. Notice that our SCT does not required any additional external connections (i.e., I/Os or pads). All of its signals are connected to existing wires of the target circuit. The invasive portion of our attack is the insertion of new cells and routing wires. Thus, the original logic of the target circuit remains unaltered. 
  
  The DV is responsible for dividing the frequency, as the name suggests. This feature is interesting for two reasons: there are scenarios when the attacker wants to slow down the speed at which bits are leaked, thus giving him/her control over the amount of time the attack will take. In other scenarios, the crypto core operates at a frequency that the trojan controller cannot match, so to avoid timing violations in the HT itself, clock dividing proves useful. Thus, the clock\_sct signal is either connected directly to the system\_clock or to the DV. The TC is responsible for enabling the RO and for connecting the leaking bits to the RO. The RO starts running when the enable signal is asserted; its operating frequency is controlled by the select signals S0 and S1 (see Fig. \ref{fig:ro_example}).
 
 To reduce the detection probability and increase the attack's feasibility, the SCT has to be customized for each targeted circuit. Therefore, the SCT is designed with size and power constraints. The size constraint is set as a percentage of the target circuit size. The power constraint is a percentage of the target circuit's idle power. As the size and power constraints are set by analyzing the circuit's physical characteristics, the attacker has to acquire such information from the layout. According to the flow detailed in Fig. \ref{fig:trojan_ins}, the layout is inspected as follows:
  
 \begin{itemize}
     \item[] \textbf{Netlist extraction:} since the attacker only holds the layout, a gate-level netlist has to be extracted. Such effort is considered a trivial task for an expert IC designer, demonstrated in \cite{gds_to_gnetlist}.
     
     \item[] \textbf{Frequency estimation:} the attacker has to estimate the operating frequency of the target circuit by performing static timing analysis on the extracted gate-level netlist. The attacker can observe the critical path(s) and then increase/decrease the frequency as needed to make the timing slack positive but near zero\footnote{Currently, our method does not take into account multi-cycle and false paths, which can reduce the accuracy of the frequency estimation but does not prevent the attack.}.
     
     \item[] \textbf{Power analysis:} with the extracted gate-level netlist and the estimated frequency, the attacker can perform a typical power analysis. For relatively large circuits, static power can be estimated very precisely even without input vectors\footnote{For crypto cores in particular, it is a fair assumption to consider the plaintext to be randomly assigned, the adversary does not need precise vectors to estimate the (order of magnitude) of the power consumption.}. 
     
 \end{itemize}
  
  Therefore, after the layout is inspected, the attacker has acquired the estimated frequency, estimated power consumption, and exact size of the circuit (number of gates). From the inspection, the attacker is now ready to draw the constraints necessary to design the SCT. First, the RO's dynamic power can be tweaked according to the power constraint. We remind the reader that the total power consumption can be divided into static and dynamic components as in (\ref{eq:p_total}). Leakage power is the static component of power and depends mainly on the threshold voltage of the transistors. On the other hand, dynamic power depends on the circuit's activity. Consequently, leakage power is proportional to the number of cells in the circuit while the dynamic power is proportional to the operating frequency. Note that, leakage power is input independent, hence, the attacker is able to acquire the original leakage assuming he/she utilizes the same corner case as the victim. Thus, to model the amplitude steps required for the RO, we need to carefully model its dynamic power consumption. 

  \begin{empheq}[]{align}
   P_{total}   &= P_{static} + P_{dynamic} \label{eq:p_total} \\
   P_{dynamic} &= \frac{1}{2} {V_{DD}}^2  F_{sa}\sum_{i_{net}} C_{load}(i) + F_{sa}\sum_{cell_j} E(j) \label{eq:p_dyn} \\
   F_{sa}      &= 2 F_{RO} = \frac{1}{\tau_{chain} } \label{eq:tr}
  \end{empheq}
  
  Dynamic power can be calculated using (\ref{eq:p_dyn}), where $C_{load}$ is the capacitance load at the output nets, $F_{sa}$ is the switching activity factor, $V_{DD}$ is the supply voltage, and $E$ is the total energy of a cell. The switching activity factor describes how many switches will occur per second. As for the RO, since the signals are always switching, this factor is two times the RO's oscillation frequency, which can be estimated by calculating the total path delay of the ring as in (\ref{eq:tr}).
  
  Our proposed SCT consists of only standard cells from the foundry-provided library. No custom cells or custom design techniques are involved. In addition, we utilize dedicated delay cells from the library for implementing the delay branches; these cells can provide delays in the range of 50ps to 4ns each. Thus, the total path delay of the RO is estimated using (\ref{eq:total_delay}), where $\tau_{delay}$ is proportional to the number of delay cells in the active path times the delay of each cell ($\tau_{dcell}$), see (\ref{eq:td_cell}). The $\tau_{inverter}$ delay is from the inverter cells in the feedback path, described by (\ref{eq:td_inver}). The $\tau_{control}$ delay is from the logic cells that controls the active paths, described by (\ref{eq:td_control}). Since $\tau_{inverter}$ and $\tau_{control}$ are fixed for a given implementation, $\tau_{delay}$ is the knob utilized to change the frequency of oscillation dynamically. Therefore, the RO can be designed by choosing the adequate number of delay cells in each individual branch as well as the (static) number of inverter cells in the feedback path.
       
    \begin{empheq}[]{align}
     \tau_{chain} &= \tau_{delay}  +  \tau_{inverter} + \tau_{control}\label{eq:total_delay} \\
     \tau_{delay} &= (N_{D1}+N_{D2}+N_{D3}+N_{D4}) \tau_{dcell} \label{eq:td_cell} \\
     \tau_{inverter} &= N_i \tau_{invcell} + \tau_{nand}\label{eq:td_inver} \\
     \tau_{control} &= 7 \tau_{and} + 3 \tau_{or} \label{eq:td_control}
     \end{empheq}
     
  Finally, the equations above give a first-order estimation of the power profile of the RO. 
  

\subsection{Side Channel Trojan Insertion} \label{sec:trojan_insertion}

After designing the SCT, the next step is its insertion. The attacker can utilize the ECO feature provided by commercial EDA tools for inserting the SCT. This feature can be used in two scenarios to fix small bugs in finalized layouts, pre- and post-mask. For post-mask, the ECO is used to perform slight modifications in a finalized layout after its manufacturing. A special type of logic cell, called a spare cell, is utilized to enable the post-mask ECO. Spare cells are typically inserted in commercial ICs and, when needed, are instantiated by the ECO flow. Doing so can generate a new design with minimal changes in the fabrication mask set. In the case of pre-mask ECO, the finalized layout was not manufactured yet; thus, \textit{it does not require any special cell}. If necessary, the ECO re-purposes empty spaces to add new cells to the layout.

The main reason for utilizing the pre-mask ECO is to avoid the time-consuming re-implementation of a design. In addition, the pre-mask ECO is a one-time operation. Therefore, this is precisely what the attacker wishes when inserting any sort of additional logic, an automated and fast manner for modifying the target layout. Hence, our attack leverages the pre-mask ECO for inserting the SCT.
 
 For the SCT insertion via ECO, an attacker can achieve his/her goal without utilizing spare cells. Since we previously established that the attacker can discern any gate in a layout, the attacker can replace both filler and spare cells for his malicious logic. Contrarily to spare cells, every layout of a digital circuit has filler cells. During placement, EDA tools have to spread the standard cells to assure routability, thus mandatorily leaving gaps between cells. For more details about the relationship between placement density and HT insertion, we direct the reader to \cite{Trippel2020}.


 According to Fig. \ref{fig:trojan_ins}, the ECO flow is the last step for the SCT insertion. After the ECO, the attacker has to perform timing sign-off to guarantee that the performance of the victim's design was not disturbed. The SCT insertion is not likely to perturb the target's performance; it is only connected to a register (crypto key storage) and some control signals, adding a small capacitance load to them. Besides, the coupling capacitance inserted by the additional routing wires is minimal due to the SCT's lightweight characteristic and the inherent goal of the ECO flow: \emph{not to disturb the existing logic}. However, even if unlikely, the addition of the SCT could hinder the target performance. In that case, it means that the size constraint used for designing the SCT was inappropriate. Since the ECO makes this type of attack relatively fast, the attacker can try different SCT architectures, and constraints, until it fits the target.
 
 The attacker also has to check whether the SCT itself has timing violations. If so, the optional clock divider must be included to slow the SCT clock (w.r.t. the system clock). Every division by two requires one additional D-type flip-flop. 
    
\section{Implementation and Simulation Results} \label{sec:results}
 
 In this section, we demonstrate our methodology and justify our chosen target designs, i.e., the crypto cores that we are going to insert our trojans on. The first part of the experiments are from simulations done using industry-grade EDA tools. After validating the methodology through simulation, the next step is the demonstration of our ASIC prototype, discussed in Section~\ref{sec:testchip}.

\subsection{Targets and Conditions}

 For our experimental investigation, we have utilized AES-128 and Present (PST) \cite{pst_cipher} crypto cores with $N_{key}=128$ and $N_{key}=80$, respectively. The AES crypto core was chosen due to its standardized status and popularity, while PST was chosen due to its lightweight characteristic \cite{present_on_ASICs}.

 To allow the analysis of SCT insertion for both AES and PST cores regarding changes in \textit{frequency} and \textit{placement density}, the combination of these variables is explored in Table~\ref{tab:def_corevars}. In the column titled `Acronym', we define the terminology used for referring to the many variants of the cores. Results from physical synthesis of the considered targets are presented in Table~\ref{tab:cores_phys_res}. A 65nm commercial CMOS technology was utilized to exercise very challenging placement densities (e.g., 75\% for AES\_LFHD) and frequencies (e.g., 0.95GHz for PST\_HFLD). The values reported are for typical process corner (TT) and a nominal temperature of 25$^\circ$C.

    \begin{table}[htbp]
        \centering
        \caption{Naming convention for the crypto cores regarding their frequency and placement density}
\begin{tabular}{cccc}
\hline
\multicolumn{1}{c}{\textbf{Core}} & \textbf{\begin{tabular}[c]{@{}c@{}}Frequency\end{tabular}} & \textbf{\begin{tabular}[c]{@{}c@{}}Placement density \end{tabular}} & \multicolumn{1}{c}{\textbf{Acronym}} \\ \hline
\multirow{4}{*}{AES} & \multirow{2}{*}{Low} & Low & \textit{AES\_LFLD} \\ \cline{3-4} 
 &  & High & \textit{AES\_LFHD} \\ \cline{2-4} 
 & \multirow{2}{*}{High} & Low & \textit{AES\_HFLD} \\ \cline{3-4} 
 &  & High & \textit{AES\_HFHD} \\ \hline
\multirow{4}{*}{PST} & \multirow{2}{*}{Low} & Low & \textit{PST\_LFLD} \\ \cline{3-4} 
 &  & High & \textit{PST\_LFHD} \\ \cline{2-4} 
 & \multirow{2}{*}{High} & Low & \textit{PST\_HFLD} \\ \cline{3-4} 
 &  & High & \textit{PST\_HFHD} \\ \hline
\end{tabular}
        \label{tab:def_corevars}
    \end{table}

\subsection{SCT Design Results}

    \begin{table*}[htbp]
    \centering
    \caption{Physical synthesis results for our considered targets, before and after trojan insertion.}
    \begin{tabular}{p{1.3cm}p{1.2cm}|p{1.0cm}p{1cm}p{.5cm}p{1.5cm}|p{1.0cm}p{1cm}p{.5cm}p{1.5cm}|p{1.4cm}}
    \hline 
           &    & \multicolumn{4}{c|}{\textbf{Before SCT insertion}} & \multicolumn{4}{c|}{\textbf{After SCT insertion}} \\
     \textbf{Core} & \textbf{Frequency (MHz)} & \textbf{Density (\%) } & \textbf{Leakage ($\mu W$)} & \textbf{CT ($\mu W$)} &   \textbf{Total Power ($\mu W$)}&  \textbf{Density (\%) } &  \textbf{Leakage ($\mu W$)} & \textbf{CT ($\mu W$)} & \textbf{Total Power ($\mu W$)} & \textbf{$\Delta$ Density (\%)}  \\
     \hline 
        AES\_LFLD & 100    &  61  & 77.4  & 115.2  & 1670   &  63.45  & 80    & 115.8 & 1720 & +2.45 \\
        AES\_LFHD & 100    &  75  & 75.8  & 116.7  & 1660   &  78.20  & 79    & 117.6 & 1720 & +3.20 \\
        AES\_HFLD & 1000   &  58  & 1048  & 1228   & 22800  &  59.37  & 1052  & 1238  & 23015 & +1.37 \\
        AES\_HFHD & 1000   &  72  & 1036  & 1241   & 22610  &  73.02  & 1040  & 1252  & 22830 & +1.02 \\
        PST\_LFLD & 95     &  53  & 14.13 & 32.05  & 371.3 &  67.33  & 20.71 & 34.75 & 483.4 & +14.33 \\
        PST\_LFHD & 95     &  70  & 14.09 & 31.89  & 371.2 &  82.05  & 17.72 & 32.85 & 428.5  & +12.05 \\
        PST\_HFLD & 950    &  52  & 34.02 & 325.30 & 3744   &  60.89  & 36.85 & 338.1 & 4022  & +8.89\\
        PST\_HFHD & 950    &  69  & 34.13 & 329.10 & 3785   &  80.26  & 36.96 & 341.5 & 4015 & +11.26 \\
     \hline 
    \end{tabular}
    \label{tab:cores_phys_res}
    \\[-10pt]
\end{table*}
 
 Initially, all studied cores were physically synthesized for the placement and frequency conditions set above. These results are obtained from an industry-grade physical synthesis tool and are reported as pre-ECO results in Table \ref{tab:cores_phys_res} (``Before SCT insertion''). As we assume the attacker has no means to stop the clock delivery to the entire circuit, we have included the clock tree (CT) power in our results, as it has to be accounted for in the SCT power constraint. Notice how the CT power is significant compared to the targets' leakage power, even for the LF variants. Different SCTs were designed for each target by setting a power budget for the SCTs to be 10\% of the sum of leakage and CT power. Since our leaking scheme does not require a high signal-to-noise ratio (SNR) to operate; we argue the 10\% margin is a good trade of capability of leaking the bits and stealthiness. Importantly, this is \textbf{not} a limitation of the methodology; an attacker can pick any other threshold and still design the SCT accordingly.
   
\begin{figure}[tbp]
     \centering
     \includegraphics[width=0.7\linewidth]{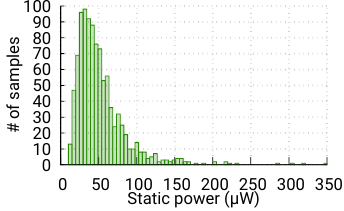} 
     \caption{PST\_HFLD static power histogram, 10K MC samples (from \cite{trojan_iscas}).}
     \label{fig:MCsimulation}
 \end{figure}

 With the goal of obtaining a better understanding of the static power consumption of the cores, we performed SPICE Monte Carlo (MC) simulations using a commercial circuit simulator. The MC simulation was performed for 10000 samples, varying only the process characteristics, with the temperature fixed to 25$^{\circ}$C. Fig.~\ref{fig:MCsimulation} depicts the static power distribution of the PST\_HFLD core. As expected, there are obvious outliers. Nevertheless, the distribution average matches the value reported in Table \ref{tab:cores_phys_res} for the typical corner. We will later show that the SCT is implemented in the very same region of the IC as the target, therefore we can also expect the same shifts in power due to process variation. This is an important observation: if a given die falls closer to the FF corner than to TT, it will be faster and more power hungry -- but so will be the trojan, nearly at the same rate. Therefore, we argue that we can safely utilize the nominal power values reported in Table \ref{tab:cores_phys_res} for RO design, regardless of the quality of the fabricated silicon.

 Once the power constraint has been established, the attacker can proceed to estimate the multiple operating frequencies of the RO (and their associated power values that effectively leak the key). For this goal, we have taken each of our SCTs and performed SPICE simulations. The oscillation frequency and power consumption of the ROs are reported in Table \ref{tab:ro_impl_res}, where each RO has been termed with a ``DXIY'' suffix, where X and Y represent the amount of delay and inverter cells, respectively. Notice how we do not differentiate density in the results reported in Table \ref{tab:ro_impl_res}: either the trojan fits or it does not. The SCT design is nearly agnostic to placement density; this is the reason why the table contains four entries instead of eight.

    \begin{table}[tbp]
    \centering
    \caption{RO operating frequency and power consumption for four variants of AES and PST.}
    \begin{tabular}{p{0.85cm}p{1.25cm}p{1.1cm}p{1.1cm}p{1.1cm}p{1.1cm}}
    \hline 
        \textbf{Target} & \textbf{RO} & \multicolumn{4}{c}{\textbf{Power \& Frequency ($\mu$W \& MHz)}} \\
        \multicolumn{2}{l}{\textbf{core}} & \textbf{S=00} & \textbf{S=01} & \textbf{S=10} & \textbf{S=11}  \\
    \hline
      AES\_LF     & $RO_{D6I10}$  & 19@65  & 17@45  & 15@34  & 13@20 \\
      AES\_HF     & $RO_{D10I10}$ & 198@551 & 182@483 & 161@390 & 140@300 \\
      PST\_LF & $RO_{D6I4}$   & 16@112 & 11@58  & 10@39  & 8@20 \\
      PST\_HF & $RO_{D8I10}$  & 42@79  & 36@61  & 31@46  & 26@31 \\
    \hline 
    \end{tabular}
    
    \label{tab:ro_impl_res}
\end{table}

\begin{figure}
    \centering
    \includegraphics[width=0.95\linewidth]{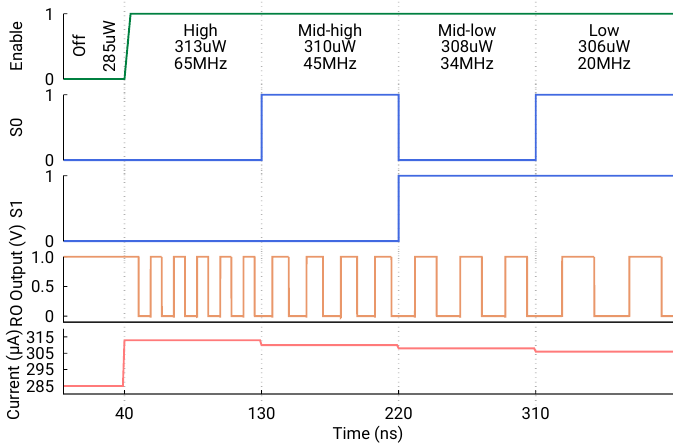}
    \caption{Post-layout simulation of our SCT architecture in Cadence Spectre. The target design is AES\_LFHD and the Trojan payload is configured as $RO_{D6I10}$ (modified from \cite{trojan_iscas}).}
    \label{fig:ro_example}
\end{figure}

 To help the reader better visualize the operation of the SCT, the illustration in Fig. \ref{fig:ro_example} displays a SPICE simulation of the SCT using the AES\_LFHD target as an example. The set of leaked keys in the image is \{00-01-10-11\}. The results for the RO are from a SPICE-level simulation with parasitics, and the total power of the AES\_LFHD is estimated from physical synthesis.

 We also highlight an extreme case in the $RO_{D6I4}$ which targets the PST\_LF core. For instance, both PST high frequency versions, PST\_HFLD and PST\_HFHD, are 2.55 and 2.6 times larger than their low frequency counterparts, respectively. Here, the SCT alone represents about 10\% of the size of the PST\_LF core. Since area and leakage have a linear dependency, the SCT's leakage already is about 10\% of the target's leakage.  Hence, the power constraint is violated. However, this extreme example assumes the entire IC consists of a single PST core. For a large system-on-chip containing multiple cores, the power budget for designing the SCT would be much more forgiving.

\begin{figure}[tb]
        \centering
        \includegraphics[width=0.95\linewidth]{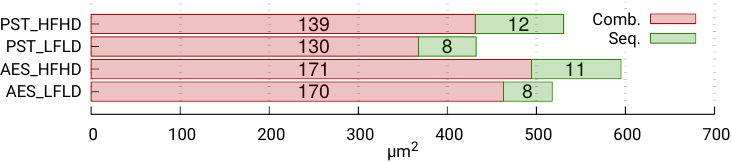}
        \caption{Comparison of area and number of cells between SCTs (from \cite{trojan_iscas}).}
        \label{fig:phys_trojan}
    \end{figure}

 Alongside the custom-simulated ROs, the SCTs are synthesized for each $N_{key}$ and at the same clock frequency as of the target. Exclusively for the HF targets, we added the DV block to ensure the synchronous portion of the SCT does not violate timing. For AES\_HF, the system clock was divided by eight while for PST\_HF it was divided by sixteen. The characteristic of the SCTs are illustrated in Fig. \ref{fig:phys_trojan}, where we show the area and number of cells for each SCT.
   
   \subsection{SCT Insertion Results}

After designing the RO and synthesizing the remainder of the SCT logic, the attacker is ready to perform the insertion via ECO. Here, \textit{we assume the role of the attacker} and utilize the post-route layout for extracting the gate-level netlist from the targets and performing the insertion methodology described in Fig.~\ref{fig:trojan_ins}. In our experiments, we complete the ECO flow in a single run, i.e., calling the ECO command a single time.

The results for SCT insertion are described on the right side of Table \ref{tab:cores_phys_res} (`After SCT insertion'). For all considered scenarios, the ECO flow was capable of placing and routing the SCT successfully, even for extremely dense layouts. Considering that high cell density implies fewer routing resources, we verified that the ECO flow purposefully utilizes the least congested metal layers. This trend is noticeable in Table \ref{tab:pst_routing}\footnote{In our considered metal stack, M1 cannot be used for signal routing. For this reason, M1 is not shown. Similarly, M8/M9 are reserved for power distribution.}, where the routing length per metal layer is reported for the PST\_HFHD target. Notice the significant increase in metals M5, M6, and M7. Also, note that the lower metal layers are more closely associated with the circuit performance \cite{relaxedbeol}, so overheads in M5 and above are unlikely to affect critical paths.

  \begin{table}[htbp]
    \centering
    \caption{Routing length per metal for the PST\_HFHD implementation, pre- and post-ECO.}
    \begin{tabular}{ccc}
    \hline  & \multicolumn{2}{c}{Wirelength ($\mu m$)} \\
        Metal layer & pre-ECO & post-ECO \\
         \hline
         M2  & 5568 & 5759 \\
         M3  & 7036 & 8332 \\
         M4  & 4580 & 6223 \\
         M5  & 3182 & 6417 \\
         M6  & 2528 & 5283 \\
         M7  & -    & 706 \\
         \hline 
    \end{tabular}
    \label{tab:pst_routing}
\end{table}

\begin{figure*}[t]
    \centering
    \includegraphics[width=0.8\linewidth]{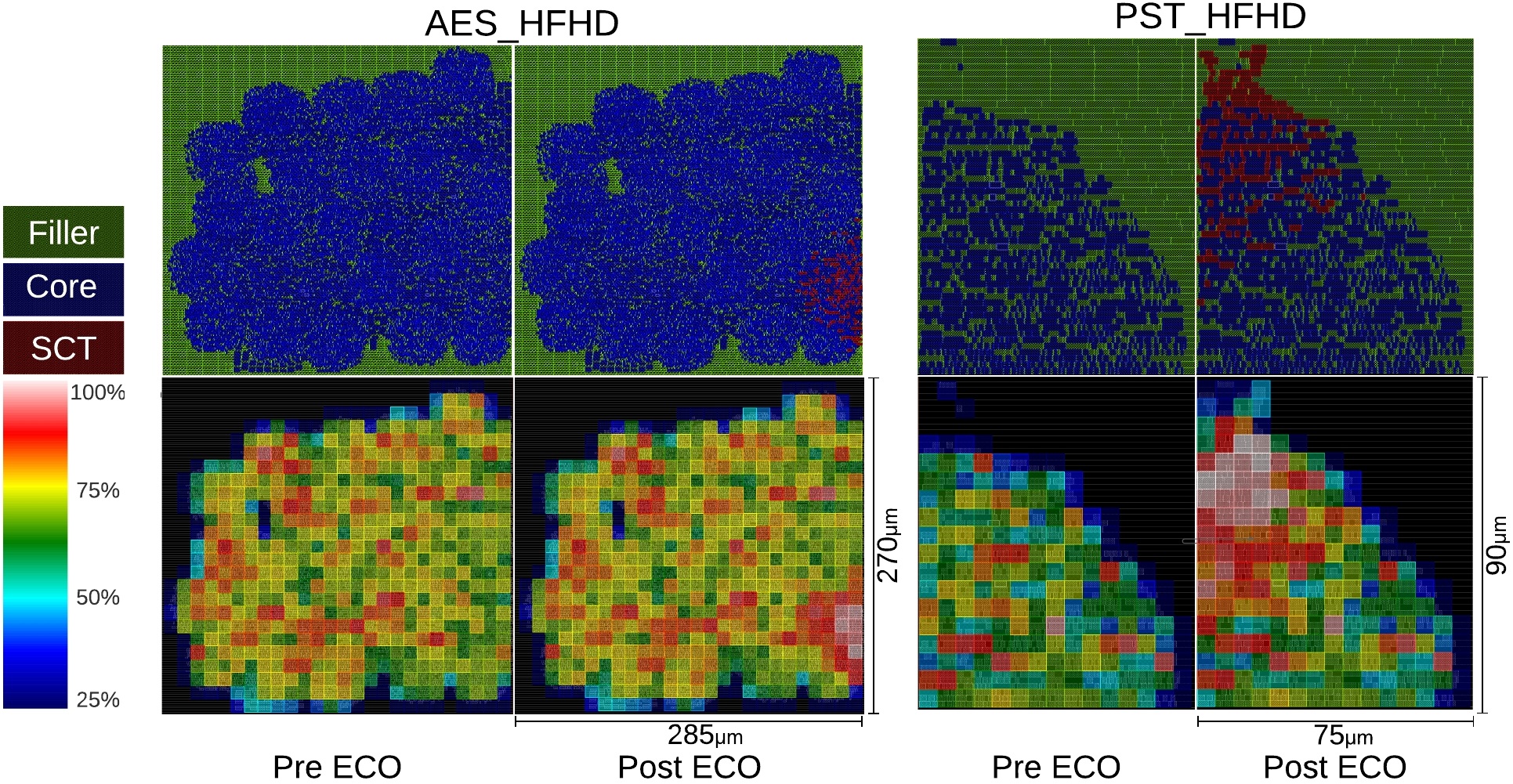}
    \caption{Placement view (top panels) and density map (bottom panels) of the AES\_HFHD and PST\_HFHD cores, before and after SCT insertion via ECO (modified from \cite{trojan_iscas}).}
    \label{fig:pst_den_place_comp}
\end{figure*}

 We also provide a visual comparison of the density increase for the AES\_HFHD and PST\_HFHD SCTs in the bottom part of Fig. \ref{fig:pst_den_place_comp}. Note that the placement of the targets (top part of Fig. \ref{fig:pst_den_place_comp}) was kept identical and only filler cells were removed for the SCT insertion via ECO. This is a key finding of our work and confirms the feasibility of the attack.

 Besides being able to insert the SCT, the ECO flow also has to preserve the performance of the target circuit. The additional malicious logic increases the load on the paths to which the SCT is connected, and, in general, the SCT routing could increase the coupling capacitance to adjacent paths. Thus, the impact on the target performance due to the SCT insertion is related to the number of connections between them and the increase in density. For the AES implementations, the addition of the SCT increased their total density by a small margin. On the other hand, for the PST cores, the SCT represents a large portion of the total circuit area. This is illustrated in the bottom part of  Fig. \ref{fig:pst_den_place_comp}, where the density map of the PST\_HFHD and AES\_HFHD layouts are shown. 
 
\begin{figure}[tb]
    \centering
    \includegraphics[width=0.95\linewidth]{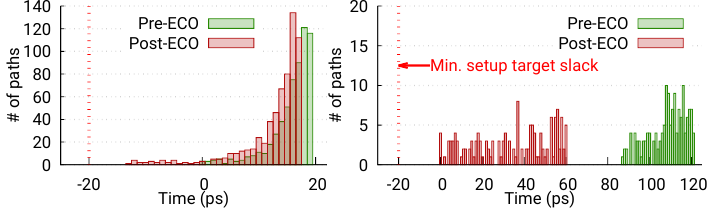}
    \caption{Pre- and post-ECO setup timing slack comparison of AES\_HFHD (right) and PST\_HFHD (left) (from \cite{trojan_iscas}). }
    \label{fig:timing_eco}
\end{figure}

 The impact on the performance of the AES\_HFHD and PST\_HFHD cores is illustrated in Fig. \ref{fig:timing_eco}, where we contrast the pre- and post-ECO timing slack. These results show that the impact is greater on the PST\_HFHD implementation, which is explained by the high density increase reported in Table \ref{tab:cores_phys_res}. One can appreciate how the red bars in Fig. \ref{fig:timing_eco} are shifted to the left (w.r.t. the green bars). However, this shift was not sufficient to degrade the performance of the cores. In particular for the PST core, the ECO engine completed successfully by using some of the safety margin (20ps) we applied to all our paths. This safety margin is small and compatible with industry practices, where often margins in the range of tens of picoseconds are utilized. Thus, in terms of performance, our attack appears to be adequate for realistic commercial designs.  
  
 Furthermore, considering that the reported timing slack results are for implementations with a challenging operating frequency, we argue that our proposed methodology is not only capable of inserting an SCT in a high density target, but also of keeping the target performance as close as possible to the original, regardless of the frequency of operation. Thus, for an attacker, inserting malicious logic by repurposing filler/spare cells with the help of an ECO feature is more than adequate, it is almost ideal. First, this methodology has an area overhead of 0\% because we utilize space that is otherwise unused. Therefore, a slight increase in density is the only measurable ``overhead''. Second, the performance of the target is very likely to remain the same. Third, the runtime is only a fraction of other methods for inserting malicious logic, such as re-implementation. In the following section, we discuss these ECO characteristics in detail. For further details of the capabilities of the ECO and a rich discussion of insertion success, we direct the reader to \cite{iccad_rtrojan}.

\section{Testchip Design and Validation} \label{sec:testchip}

 In this section, we present our fabricated ASIC prototype and its many details. Initially, we present the chip architecture and its functionality. 
 
 \subsection{ASIC Prototype Architecture}

 Our main goal while designing a silicon proof of concept for our methodology is to demonstrate the malicious potential of the ECO flow. For this purpose, we developed a full framework for performing a fabrication-type attack (see Fig.~\ref{fig:trojan_ins}). Our proposed SCTs are carefully crafted in order to stress test the ECO flow and its limitations: the chosen circuits are synthesized for their maximum frequency and utilizing challenges densities, making the SCT insertion even more challenging. Our framework includes all steps necessary for assessing the GDSII database, designing a hardware trojan, and inserting it in a finalized layout.
 
\begin{figure}[htb]
    \centering
    \includegraphics[width=0.75\linewidth]{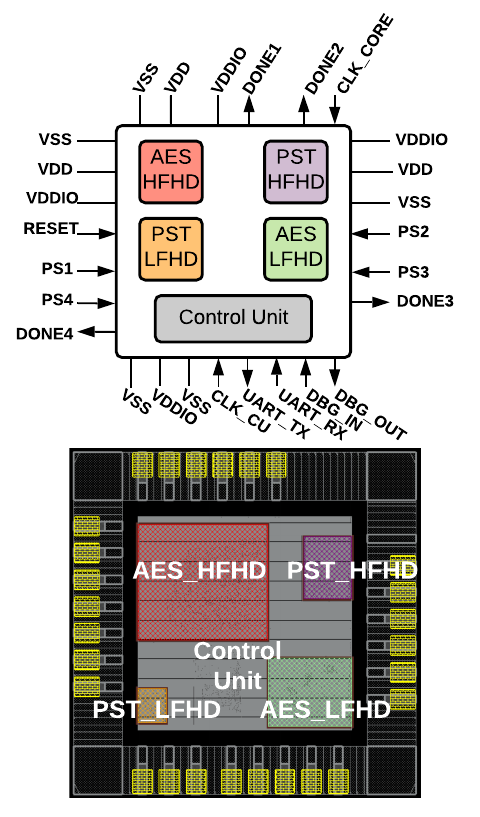}
    \caption{Top level diagram (top panel) and floorplan of our testchip (bottom panel).}
    \label{fig:toperson}
    \vspace{-0.5cm}
\end{figure}

 As discussed in the previous section, our targets are AES and PST crypto cores. For our ASIC prototype, we chose 4 of the 8 versions of the crypto cores described in Section~\ref{sec:results}. These versions are the highest density ones (AES\_LFHD, AES\_HFHD, PST\_LFHD, and PST\_HDHD), purposefully selected for the difficulty in manipulating a dense layout. The top level architecture and the floorplan of our ASIC design are depicted in Fig.~\ref{fig:toperson}. 

 Our chip contains the four chosen crypto cores and a control unit for handling the data traffic in and out of the chip. This control unit has a UART-like communication protocol and a register bank to store the plaintext, the cryptokey, and the ciphertext. We note that the plaintext and the cryptokey can be programmed externally via UART. For communicating with the control unit, the signals UART\_TX and UART\_RX are utilized. The signals DONE\_1, DONE\_2, DONE\_3, and, DONE\_4 indicate the end of a cryptographic operation for AES\_HFHD, AES\_LFHD, PST\_HFHD, and PST\_LFHD, respectively. These signals are exposed as primary outputs only for debug reasons, their presence is not required for the attack. Internally, these same signals are the triggers for the SCTs. 
 
  \begin{table}[b]
    \centering
    \caption{Power domains, clock, average total power, and, leakage across the samples tested.}
    \begin{tabular}{p{1.1cm}p{1.2cm}p{0.8cm}p{1.5cm}p{1.5cm}}
        \hline
         \textbf{Block} & \textbf{Clock} & \textbf{Switch Signal} &  \textbf{Leakage ($\mu$W)} & \textbf{Total Power ($\mu$W)}\\
        \hline
          Control Unit & CLK\_CU @1MHz  & Always on   &  46.69$\pm$4.75 & - \\
          AES\_HFHD    & CLK\_CORE @1GHz  & PS1  & 743.79$\pm$108.07 & 101160$\pm$10781 \\
          AES\_LFHD    & CLK\_CORE @100MHz & PS2  & 131.57$\pm$10.35 & 3139.32$\pm$85.38 \\
          PST\_HFHD    & CLK\_CORE @950MHz & PS3  & 80.75$\pm$7.82 & 9661.3$\pm$758.52 \\
          PST\_LFHD    & CLK\_CORE @95MHz & PS4  &  74.35$\pm$6.84 & 868.56$\pm$57.90 \\
         \hline
    \end{tabular}
    \label{tab:pd_leakage}
\end{table}
 
 Our architecture has 5 clocks domains: the ``always-on'' clock is delivered by the signal CLK\_CU, and the other 4 domains are connected to the signal CLK\_CORE (which assumes 4 different frequencies). For switching the cores on/off selectively, the signals PS1, PS2, PS3 and, PS4 are utilized as described in Tab.~\ref{tab:pd_leakage}. Both signals DBG\_IN and DBG\_OUT are used for debugging purposes only. The power-ground scheme has 2 power sources (VDD and VDDIO) and a common ground (VSS): VDD supplies the core cells with a nominal voltage of 1.0V and VDDIO supplies the IO cells with a nominal voltage of 3.0V.  
  
 For manufacturing the chip, we have utilized a commercial 65nm technology (the exact same technology utilized in the previous section). We also made use of three standard cell flavors (LVT, SVT, and HVT) and power switch IP for isolating power domains. Our idea in utilizing multiple voltage threshold cells is to bring our implementations in line with industry practices, thus adding another degree of realism to our attack. The power switches were utilized to create a power domain for each crypto core, making it possible to enable one core at a time on the same chip. Implementing the crypto cores with the possibility of total shut-down is extremely valuable for evaluating our attack, because we are only reading the power that come from the enabled core. However, in our chip, the control unit is on an ``always-on'' domain, thus, this portion of the circuit is always enabled. Nonetheless, this characteristic later did not affect our tests or measurements in a negative manner. The power domain information and the related switch signals are described in Tab.~\ref{tab:pd_leakage}. 

 The ideal ROs designed in the previous section (see Tab.~\ref{tab:ro_impl_res}) only consider the leakage from the crypto core alone. In our ASIC prototype, we have extra components that compete with the leakage of the currently enabled core  -- even if here we assume that only one core is on at a time. Therefore, the ROs require small adjustments to accommodate the extra competing leakage, which is a trivial exercise: an attacker can create a database of SCT architectures for known targets and apply small shifts to them by modulating the number of inverters or delay cells in the RO. The newly adjusted ROs for the ASIC prototype are described in Tab.\ref{tab:ro_chip}. These results are from detailed SPICE-level simulations. 

\begin{figure}[htb]
    \centering
    \includegraphics[width=0.95\linewidth]{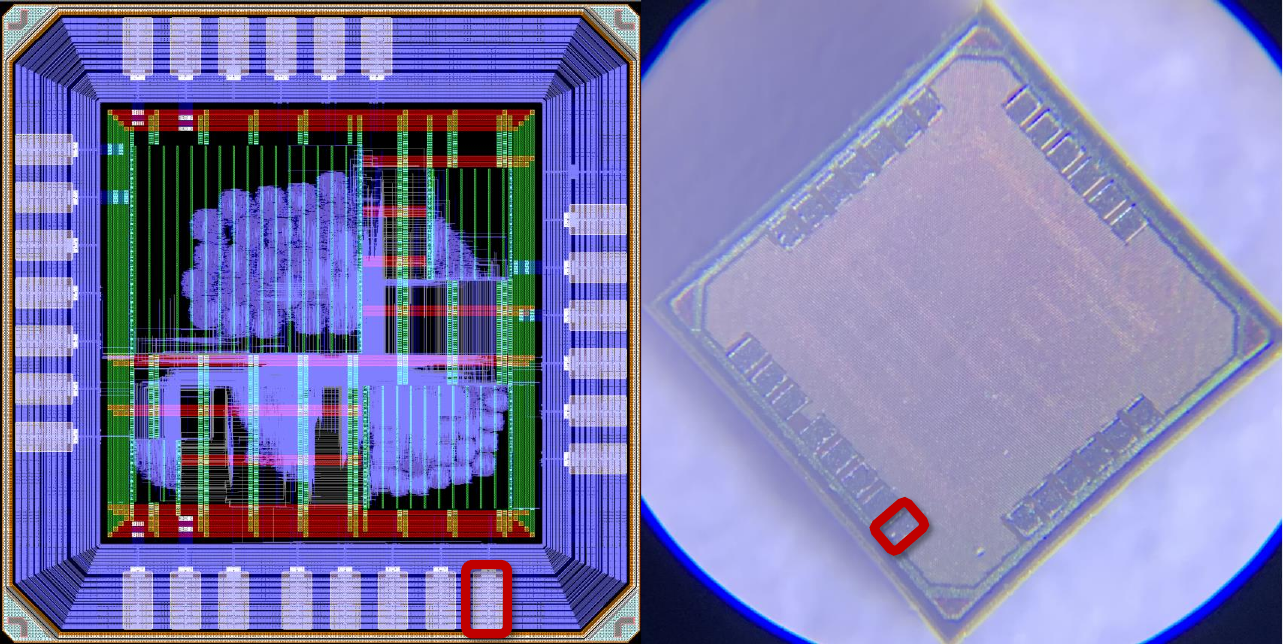}
    \caption{Our bare die (right) and its layout (left). The lower-right corner is identified by the highlighted pin.}
    \label{fig:chipao}
\end{figure}

 Our chip was designed in November 2020, fabricated at a partner foundry in March 2021, and bench tests were started in July 2021. Our bare die and its layout are contrasted in Fig. \ref{fig:chipao}. For the validation of the design, we have 25 packaged samples of the chip. All packaged samples were confirmed to be 100\% functional.

   \begin{table}[tbp]
    \centering
    \caption{RO operating frequency and power consumption for each crypto core of the ASIC prototype.}
    \begin{tabular}{p{1.1cm}p{1.25cm}p{1.0cm}p{1.0cm}p{1.0cm}p{1.0cm}}
    \hline 
        \textbf{Target} & \textbf{RO} & \multicolumn{4}{c}{\textbf{Power \& RO Frequency ($\mu$W \& MHz)}} \\
        \multicolumn{2}{l}{\textbf{core}} & \textbf{S=00} & \textbf{S=01} & \textbf{S=10} & \textbf{S=11}  \\
    \hline
      AES\_LFHD  &  $RO_{D8I14}$ & 32@90    & 27@61   & 23@46   & 20@31 \\
      AES\_HFHD  & $RO_{D12I14}$ & 249@551  & 227@483 & 198@390 & 169@300 \\
      PST\_LFHD  & $RO_{D8I6}$   & 22@169   & 19@90   & 16@46   & 13@21 \\
      PST\_HFHD  & $RO_{D10I10}$  & 30@90    & 24@60   & 20@37   & 17@19 \\
    \hline 
    \end{tabular}
     
    \label{tab:ro_chip}
\end{table}

 %
 %
 %
 %

    \begin{figure}[tb]
        \centering
        \includegraphics[width=0.95\linewidth]{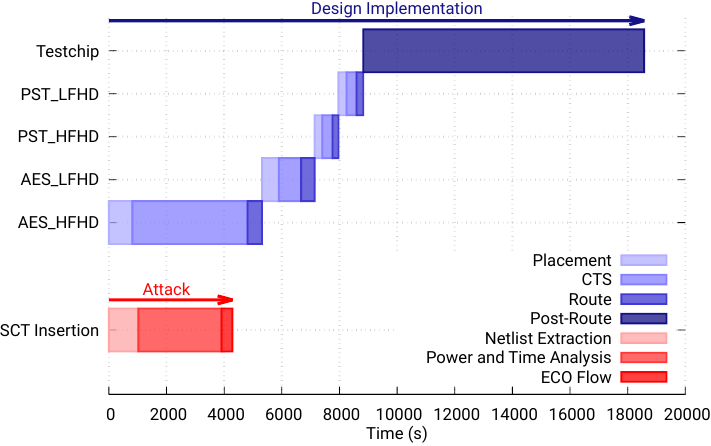}
        \caption{Physical implementation execution time (s) for each step of the flow, and, execution time (s) for inserting the SCT in each implemented crypto core.}
        \label{fig:runtime}
        \vspace{-0.5cm}
    \end{figure}

\subsection{SCT Insertion}
    
 In our framework, for fully inserting the SCT into a layout, the attacker has to inspect the layout, extract the netlist, estimate the operating frequency, estimate the power consumption, modify the netlist, and finally, insert the SCT utilizing the ECO flow. The time necessary to inspect the layout in order to find the security-critical nodes depends on the expertise of the attacker, which makes it very difficult to estimate. On the other hand, the other steps can have their runtime measured. These times are depicted in red in Fig.~\ref{fig:runtime} and are contrasted with the total time required for the physical implementation of the original design, shown in blue. In our case, the required time for implementing the original cores and building the top-level layout is about 5 hours and 9 minutes. The most time-intensive tasks are the clock-tree synthesis and the post-route optimizations. For our testchip, the post-route step also includes the assembling of the cores, the top-level clock-tree synthesis, the top-level routing, and a chip-level design rule check. The netlist extraction was executed in a server with an \textit{Intel Xeon} 5122 CPU @ 3.6GHz and 96GB of RAM, while all other tasks were executed in a server with an \textit{Intel Xeon} X5690 @ 3.47Ghz and 768GB of RAM. When multithreading was supported for a given task, the number of concurrent CPUs in use was set to 8. All execution times were measured 5 times and the presented values are the average execution time.
 
 For the attack, the first task is extracting the netlist from the layout, which is done in 17 minutes. Then, the estimation of the operating frequency and power consumption is done in 48 minutes, which includes full parasitic extraction, static timing analysis, and a power analysis -- all utilizing the highest effort available. The static timing analysis was done utilizing two corners and the power analysis only one. Differently of a chip sign-off task executed by a designer, where all corners are considered, the attacker can make use of a couple of corners only for a representative evaluation of the IC operation. Inserting the SCT is done individually for each core, and the total time for inserting all four SCTs is only 6 minutes. Thus, the entire attack for this design requires 1 hours and 11 minutes. Had an attacker decided to modify the layout by re-implementing the extracted netlist, the total time of the attacks jumps to 6 hours and 21 minutes. Therefore, by using an ECO flow, not only the original design remains untouched -- i.e., the cells' placement and routing remain the same after the attack -- the attacker can drastically reduce the attack time. This is true for both SCTs and functional trojans.
 
 \textbf{\textit{For large ICs, how can an adversary insert a HT in a limited time frame?}} Or, in other words, how does the attack time scale with the complexity of the IC? To answer this question, we ought to look at the individual steps of the attack. First, for the netlist extraction, there are no alternative routes for avoiding/bypassing this specific task in the EDA tools. However, the execution time scales fairly with the size of the circuit.
 
 On the other hand, the timing and power analysis can become overwhelming depending on the circuit size. Fortunately for the attacker, the runtime for these tasks can be reduced by utilizing different effort levels and/or by reducing the number of corners utilized. Another option is utilizing a wire-load model instead of a more precise RC model for the parasitic extraction. All of those alternatives would change the quality of the analysis, e.g., the static timing analysis in low effort with only the best-case corner will report a very optimistic timing requirement for setup (i.e., a higher operating frequency). From the attacker's point of view, an optimistic timing analysis would lead to a higher power budget, making the SCT less stealthy but more likely to succeed. An overall picture of the trade-offs between the analysis quality, the SCT insertion runtime, SCT stealthiness, and the attack success rate is provided in Tab.~\ref{tab:rt_large_cir}. Here we clarify that the number of corners considered for the timing analysis is always two (BC/WC analysis). For the power analysis, only one corner is taken into account. The RC model utilized for the low effort is the wire-load model; other efforts utilize more precise models.
 
 It must also be highlighted that the ECO insertion time does not scale with chip size: the ecoPlace and ecoRoute engines are available within a physical synthesis tool specifically for the purpose of performing local changes. 
 
 
 Thus, even for large ICs, the attacker can significantly reduce the time required for inserting the SCT if he/she is willing to compromise the trojan stealthiness or its success rate. But, most importantly, the time required for SCT insertion when utilizing our proposed framework is always a fraction of the time required for implementing the original design (comparatively shown in Fig.~\ref{fig:runtime}). This relationship does not depend entirely on the size of the circuit.
 
 As shown on the last row of Tab.~\ref{tab:rt_large_cir}, an adversary could potentially make a `cut' in the GDSII database and treat only a part of it. To some extent, this would be the equivalent of doing a block analysis instead of chip analysis. The possibility of cutting a layout, although supported by the tools, is not studied in this work for the reason that an adversary would have to execute said cut very precisely and, after the SCT insertion, merge the modified and original layout, a task that is not trivial and could potentially damage the original layout if mishandled. 
 
 \begin{table*}[htb]
     \centering
     \caption{Trade-off comparison between the SCT insertion runtime, SCT stealthiness, and the attack success rate, for different effort configurations of parasitic extraction, static timing analysis, and power analysis.}
     \begin{tabular}{cccccccc}
     \hline
        \textbf{Cut Layout} & \textbf{Parasitic Extraction} & \textbf{Static Timing Analysis} & \textbf{Power Analysis}  & \textbf{Runtime} & \textbf{SCT Stealthiness} & \textbf{Attack Success Rate} \\
         \hline
          No & Low    & Low    & Low     &  {\color{green} Short}  &  {\color{red} Weak} & {\color{orange} Medium} \\
         No & Medium & Medium & Medium  &  {\color{orange} Average} & {\color{orange} Strong}  & {\color{orange} Medium} \\
        
         No & High   & High   & High    &  {\color{red} Very Long} & {\color{green} Very Strong}  & {\color{green} Very High} \\
         
         Yes & High   & High   & High    &  {\color{green} Short} & {\color{green} Very Strong}  & {\color{green} Very High} \\
         \hline
     \end{tabular}
     \label{tab:rt_large_cir}
 \end{table*}
 
\subsection{Tests and Measurements}

\begin{figure}[htb]
    \centering
    \includegraphics[width=0.95\linewidth]{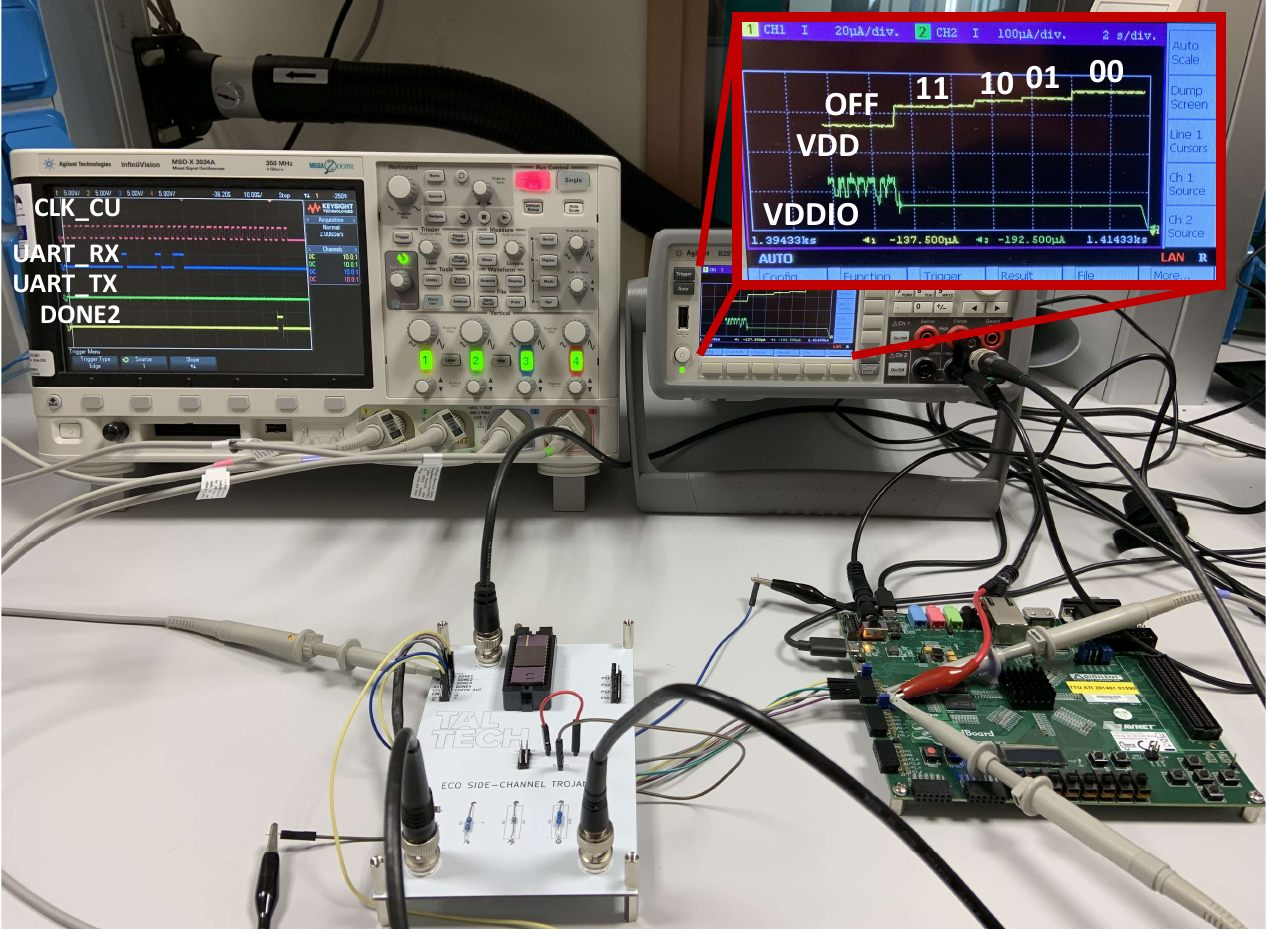}
    \caption{Setup used for bringing up the testchip. On the left side, we show the signals used for controlling the chips. On the right side, the current consumption of the chip when the RO is active.}
    \label{fig:setup}
\end{figure}

For the purpose of bench testing our ASIC prototype, we have designed a custom printed circuit board (PCB). The PCB itself only contains passive components utilized for helping with the measurements and filtering noise from the power supplies. An attacker does not need our PCB and/or test setup to mount the attack. The chip is controlled by a ZedBoard from Avnet with a Xilinx Zynq-7000 All Programmable SoC. Our complete bench test setup is shown in Fig.~\ref{fig:setup}, where we also make use of a 4-channel digital oscilloscope and a 2-channel power supply with an ammeter with pico ampere precision.

\begin{figure}[htb]
     \centering
     \includegraphics[width=0.95\linewidth]{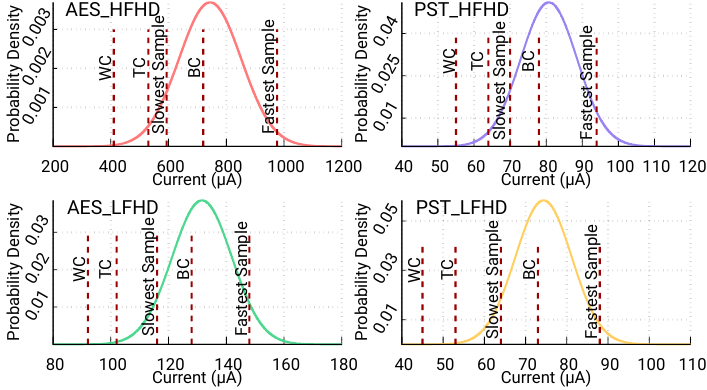}
     \caption{Leakage distribution for each crypto core contrasted with the leakage from physical synthesis report for 3 corner cases, and, the leakage of outlier samples.}
     \label{fig:dist_lkg}
 \end{figure}

 The first phase of the validation was to measure total power and leakage power from each block across all 25 samples. For this, all the primary inputs were set to ``0'', and each core was enabled one at a time by asserting its respective switch signal (see Tab.~\ref{tab:pd_leakage}). For the total power, we delivered the clock signal for each block utilizing the specified operating frequency. The results of the total power average and leakage are given in Tab. \ref{tab:pd_leakage}, and its distribution across the samples is depicted in Fig~\ref{fig:dist_lkg}. These results are in line with the expected from the power reports done during the physical implementation, as these results are contrasted in Fig.~\ref{fig:dist_lkg} for the worst, typical, and, best-case scenarios (SS-0.9V-0$^o$C, TT-1V-25$^o$C, FF-1.1V-125$^o$C, respectively). The corners provided by the vendor are for extreme cases, i.e., the best-case scenario is characterized at 125$^o$ with an over voltage of 1.1V, in our case the measurements were done at room temperature and at a nominal voltage of 1.0V. Our samples are skewed towards the best case scenario, demonstrating higher average leakage. The slowest sample is near the typical case while the fastest sample is very far from the expected best case. Thus, our samples have a high variance between them (i.e., their leakage values are very spread from the mean), with variance values of 1212, 102, 59, and 44 for the AES\_HFHD, AES\_LFHD, PST\_HFHD, and PST\_LFHD cores, respectively. 
 
 In the second phase of the experiments, we assessed the SCTs and the feasibility of the attack. This was done by the following procedure:
 
 \begin{itemize}
 
     \item A cryptokey with the 8 first bits set to ``11-10-01-00'' was programmed in the Control Unit's register bank
     
     \item A command for a single encryption was issued
     
     \item Right after the encryption is done, the chip asserts one of the ``DONE'' outputs to mark the time at which the RO starts operating
     
     \item Using only the clock signal CLK\_CORE, three bursts of clocks were sent in order to shift the cryptokey connected to the RO three times
     
     \item During the whole procedure, the current consumed by the chip is monitored 
     
 \end{itemize}
 
\begin{figure}[htb]
    \centering
    \includegraphics[width=0.95\linewidth]{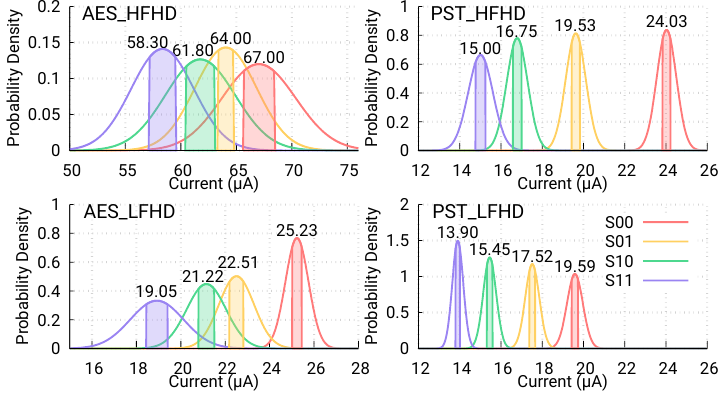}
    \caption{Power consumption ``steps'' distribution for each crypto core. The shadowed area represents the 95\% confidence interval.}
    \label{fig:ro_steps}
\end{figure}

 An example of this procedure for the AES\_LFHD core is shown in Fig. \ref{fig:setup}; the ``UART\_TX'' signal carries  the single encrypt command. As a visual aid, as soon as the ``DONE'' signal is asserted, the clock sources are turned off in this example (in Fig. \ref{fig:setup} only CLK\_CU is shown). As clearly depicted in the ammeter, there are discrete steps representing the leaked bits ``11-10-01-00'' from left to the right, respectively, as expected from the key programmed for this experiment. This was repeated 3 times for each core of each chip to confirm the behavior. The measured current values were approximated to normal distributions, as represented in Fig. \ref{fig:ro_steps}. 
 
 By comparing the RO performance from the simulations (see~Tab.\ref{tab:ro_chip}) with ASIC measurements, it is clear that the slowest ROs are performing as expected. However, the fastest RO targeting the AES\_HFHD core can only operate at a low frequency, generating a power step of about 25\% of what was expected. In this case, the ECO insertion had to spread the RO cells farther away because of the lack of empty spaces nearby (see Fig.~\ref{fig:pst_den_place_comp}). For this core, the planned power steps were in the order of 200~$\mu A$, and the actual power steps after manufacturing were in the order of 60~$\mu A$. However, the attack will still enjoy a high chance of success due to the distinct separation of the power steps, even if 95\% confidence intervals of the distributions almost overlap. We have also confirmed that the interconnect delay wire load was higher than we expected from SPICE-level simulations. For extreme cases such as the AES\_HFHD core, the attacker has to make extra considerations for implementing a RO with high power consumption, in case he/she desires high fidelity from the RO power steps. The best-case scenario is achieved for PST\_LFHD (see Fig.~\ref{fig:ro_steps}, bottom right): there is absolutely no overlapping between adjacent steps,  with very low variance, which highly increases the success rate of the attack.
 
  The experimental measurement results obtained show that the variability in the manufacturing process does not affect the effectiveness of the RO for the smaller designs (AES\_LFHD, PST\_LFHD, and PST\_HFHD), meaning that the attack can be carried out with the same probability of success, regardless of the silicon quality for a given sample. The quality of the silicon impact on the SCT performance is directly connected with its size and how the cells were placed. If the cells are heavily spread, as occurred in the AES\_HFHD, the variance of the steps is very high when compared with an SCT with a similar size (see Fig.~\ref{fig:phys_trojan}) placed with low spread, as in the PST\_HFHD. We hypothesize that the higher the physical spread between the cells that compose the SCT, the more susceptible to local variation they become. Visually, this can be seen by the width of the shadowed areas in Fig.~\ref{fig:phys_trojan}.
  
  However, we learned from our experimental results that even in a high spread scenario, the attack is successful for all implemented cores. For a single given die, there is no difficulty in differentiating the 4 possible leakage states associated with the two bits of the key. Moreover, these results make it evident that our SCT is successfully capable of creating distinct steps with a 2$\mu$A precision. Furthermore, the induced power consumption for the smaller crypto cores (PST\_LFHD) was in the 20$\mu$W range. This makes our SCT a perfect fit for targeting designs that consume very low power while maintaining a reasonable level of stealthiness.
  
\section{Discussion} \label{sec:discussion}

 For a SCT insertion framework like ours, its effectiveness can be determined by three characteristics: (1) the success rate of the attack, (2) probability of detection (i.e., its stealthiness), and, (3) feasibility of the insertion of the malicious logic during the fabrication-time attack. As we have already discussed, our SCT was successful in (1) by making the attack of leaking the cryptokey viable, i.e., the attack was fully accomplished. Nevertheless, we have not yet discussed the probability of our SCT being detected.
 
 Detecting a trojan of any kind is generally a difficult task \cite{sur_ht_det}. For SCTs, any method that relies on observing corrupted bits or any degree of incorrect computation is bound to fail -- SCTs do not alter the functionality of the device under attack. Because of the inherent opaqueness of ICs, inspecting their internal components is not trivial. Methods for inspecting ICs are separated into two classes, invasive and non-invasive. Invasive methods are generally done by delaminating the IC to reconstruct the layout layers, which leads to the destruction of the inspected sample. Our SCT is likely to be detected by an invasive method due to its size and amount of connected wires. However, these techniques are incredibly time consuming and also require costly and precise equipment. We emphasize that it is not a standard practice of the IC industry to perform this type of analysis.
 
 Differently, non-invasive methods include analyzing physical characteristics of the IC, and/or, the behavior of the IO signals (i.e., timing and state) \cite{Tehranipoor2010}. Our SCT does not disrupt any data path and our insertion methodology also does not interfere with the overall performance of the target. Thus, the probability of it being detected by analyzing the IO signals is effectively zero. Detection techniques that consider the path delay, e.g., path delay fingerprint \cite{path_delay}, would have a low probability to detect our SCT. Nonetheless, our SCT changes the overall power consumption of the target. First, the extra leakage could raise a red flag if the IC is thoroughly inspected. The chance of being detected in this type of inspection is related with the percentage of the extra leakage from the SCT. This is also true for any HT that inserts additional logic. However, if the percentage is insignificant compared with the target, the extra leakage has a high chance to be interpreted as a skew from the manufacturing process and/or imprecision of the measurements. Second, the artificial extra consumption when the SCT is triggered can also be a red flag. In this scenario, the engineer conducting the inspection would need to know the exact moment when the SCT is triggered to suspect any alteration. Specialized detection methodologies have been proposed that utilize leakage and total power as input \cite{ht_detection1, ht_detection2}. By utilizing these advanced methods of detection, our SCT could be detected due to the trigger scheme. Since our SCT is triggered after each cryptographic operation, a periodic power fluctuation would be visible. However, the trigger scheme utilized in our silicon validation can be further modulated by an attacker by creating rarer trigger conditions, making the SCT stealthier.  
 
 In our threat model, we assumed the attacker only has access to the layout and utilizes the extracted netlist for inserting the SCT. This netlist does not contain any node names, making it impossible to distinguish nodes of interest by name. Thus, the attacker has to identify the circuit functionality by inspecting the layout for `clues'. In the case of AES, this target circuit can be easily identified in a layout because of its implementation regularity, and, from there, the same holds true for the nets/registers that carry the cryptokey. On the other hand, visually locating a Present core (or any other core) would be a more challenging task. Nevertheless, visual inspection is not the only technique that can be utilized for searching for security-relevant nodes. Based on the ECO framework herein presented, the authors of \cite{iccad_rtrojan} create a framework for blindly inserting HTs. The results presented in \cite{iccad_rtrojan} further demonstrated the ECO capabilities for inserting HTs. On top of that, their framework does not require additional knowledge besides the victim's layout for locating security-relevant nodes.
 

 A clear advantage of our SCT architecture is its robustness to manufacturing process skew. As demonstrated in the previous section, even with a large difference in performance between the fastest and slowest sample (see Fig.~\ref{fig:dist_lkg}), the SCA was successful (see Fig.~\ref{fig:ro_steps}). When compared with a similar approach presented in \cite{HT_silicon}, our SCT architecture does not need any workaround to be implemented because of the performance of the technology. Even more, our flow anticipates difficulties during the SCT insertion by including the optional clock divider. However, our architecture has the disadvantage of being large in comparison with other similar SCTs. The size of our SCT is proportional to the number of bits leaked at a time and to the total number of bits intended to be leaked (i.e., the SCT is proportional to the key length). This characteristic increases the probability of detection.

  Our attack is arguably prevented by a few techniques. Split Manufacturing \cite{split_survey}, as mentioned before, is a powerful prevention technique for HT insertion overall, where the attacker has access only to the layer that contains the devices - the connections between them are hidden from the untrusted foundry. Hence, the attacker would only be able to find the nodes by visual inspection without any connection information, making the SCT insertion a `blind' effort. Another relevant technique is the insertion of dummy cells and routing wires \cite{dummy_wire, bisa_lol} with the intent to reduce the empty spaces where -- potentially -- a HT could be inserted. As demonstrated in this work, our insertion methodology overcomes incredibly high densities. Hence, these techniques would only be effective if the entire chip is populated with dummy cells and routing wires, increasing the design density above 95\%, which for new technologies is very challenging and can potentially hinder the IC performance. On top of that, dummy cells have transistors inside them, which increases the leakage of the chip proportionally to the number of additional dummy cells. Thus, the leakage overhead could be in the range of 40-50\% -- assuming that a typical design has an approximate density in the range of 60-50\%. For industrial designs, this type of technique might not be practical in terms of the potential performance loss, making the trade-off between security and power consumption not interesting for many vendors and applications. Therefore, the adoption of these techniques as a countermeasure against malicious logic is very unlikely.

 Another metric to qualify an SCT insertion attack is the total time required to perform the attack. Our threat model assumes the attack occurs in the untrusted foundry and only the layout is accessible to a rogue element. Foundries are typically working at full capacity year-round, hence, the timing window that the layout is processed to begin the manufacturing is limited. This time window is precisely the period of time in which a rogue element has to mount a fabrication-time attack. In recent SCT works that contemplate silicon validation \cite{HT_silicon, WLTrojan, sct_param}, the techniques for inserting the malicious logic are not disclosed -- making it difficult to address if the attack can be replicated in a realistic scenario.
 
 Placing and routing an SCT manually is a time-intensive task and prone to errors, even if the HT design has only a dozen of cells. Thus, the insertion of an SCT has to be automated by utilizing an EDA tool. Inserting an SCT by re-implementing the design has a significant runtime, in the case of our testchip (see Fig.~\ref{fig:runtime}) it is required at least 7 hours and 18 minutes. Replicating an entire chip without the original timing and power constraints could be very difficult, which can potentially affect the target performance, thus decreasing the stealthiness of the attack. In the case of the ECO flow, the runtime for inserting the SCT in our testchip is only 1 hours and 11 minutes. Nonetheless, as previously alluded, the ECO flow has the advantage of keeping the original design untouched which increases the stealthiness of the attack. Even more, our proposed ECO flow does not require the original power and timing constraints, an estimation can be used (see Secion~\ref{sec:trojan_design}). Consequently, our proposed ECO flow method for inserting not only SCTs, but any type of malicious logic, is arguably a superior option. Furthermore, the short runtime associated with the ECO flow makes the fabrication-time attack feasible in a realistic scenario, where the time window that a rogue engineer has for modifying the layout is (very) limited.

\section{Conclusions} \label{sec:conclusion}

 In order to steal secret information from crypto-capable ICs, a rogue element within the foundry may insert a side-channel trojan. The SCT architecture described in this work has the advantage of not violating any design specification of the target circuit, nor is any datapath obstructed. This is all possible because of the use of an ECO flow for inserting SCTs. Since this feature is readily available, adversaries may maliciously utilize it.

 Our findings and results from the validation of our ASIC prototype demonstrated the feasibility of the framework -- from the layout inspection to the actual attack. The attack was successful for all 25 samples available, successfully extracting the cryptokey via power signatures. The measurements have also demonstrated the robustness of the SCT against skews from the manufacturing process. 
 
 For our testchip, all 4 SCTs were inserted in less than two hours. Consequently, the attack would be viable in a real fabrication-time attack. As a venue for future work, we intend to improve the search of security-relevant nodes for inserting hardware trojans in order to understand if a capable adversary is able to execute a ``blind'' yet successful insertion. This line of work has already been pursued in \cite{iccad_rtrojan}, but we believe that, when armed with an ECO-based insertion engine, there are still many other ways in which a clever adversary can succeed. 

\bibliographystyle{ieeetr}
\bibliography{sd_trojan}


\begin{IEEEbiography}[{\includegraphics[width=1in,height=1.25in,clip,keepaspectratio]{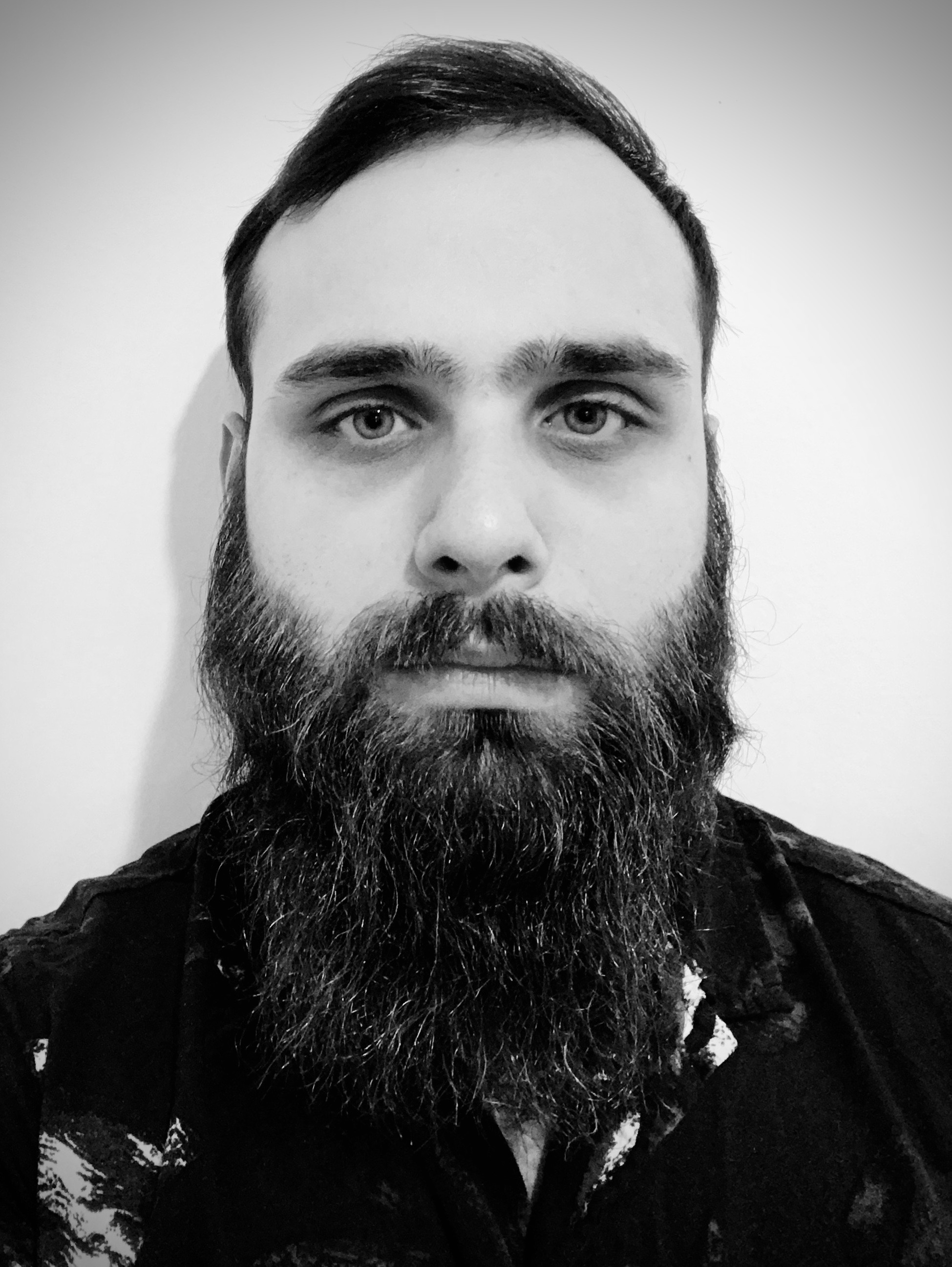}}]{Tiago D. Perez} received the M.S. degree in electric engineering from the University of Campinas, S\~ao Paulo, Brazil, in 2019. He is currently pursuing a Ph.D. degree at Tallinn University of Technology (TalTech), Tallinn, Estonia. 

From 2014 to 2019, he was a Digital Designer Engineer with Eldorado Research Institute, S\~ao Paulo, Brazil. His fields of work include digital signal processing, telecommunication systems and IC implementation. His current research interests include the study of hardware security from the point of view of digital circuit design and IC implementation.

\end{IEEEbiography}


\begin{IEEEbiography}[{\includegraphics[width=1in,height=1.25in,clip,keepaspectratio]{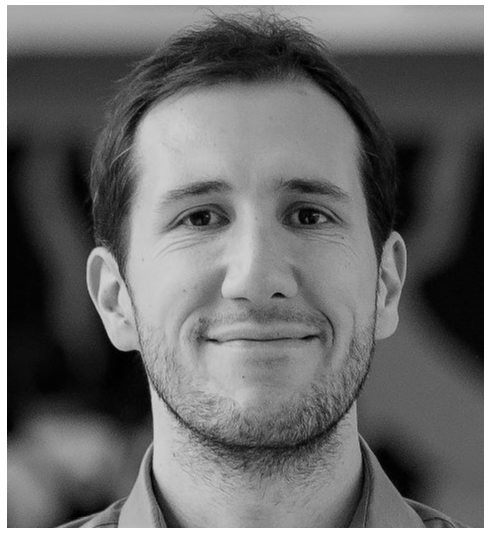}}]{Samuel Pagliarini}
(M'14) received the PhD degree from Telecom ParisTech, Paris, France, in 2013. 

He has held research positions with the University of Bristol, Bristol, UK, and with Carnegie Mellon University, Pittsburgh, PA, USA. He is currently a Professor of Hardware Security with Tallinn University of Technology (TalTech) in Tallinn, Estonia where he leads the Centre for Hardware Security. His current research interests include many facets of digital circuit design, with a focus on circuit reliability, dependability, and hardware trustworthiness.

\end{IEEEbiography}

\end{document}